\documentclass[iop]{emulateapj}
\usepackage{graphicx,epsfig,fancyhdr,rotating,amsmath} 
\usepackage{txfonts}
\usepackage{epsfig,rotate,morefloats,color}

\shorttitle{Coronal Loop Related to Active Region Transient Brightening}
\shortauthors{Gupta, Sarkar and Tripathi}
\begin{document}
\title{Observation and Modeling of Chromospheric Evaporation in a Coronal Loop Related to Active Region Transient Brightening}
\author{G.~R. Gupta\altaffilmark{1}{$^{,3}$}, Aveek Sarkar\altaffilmark{2}, Durgesh Tripathi\altaffilmark{1}}
\affil{\altaffilmark{1}Inter-University Centre for Astronomy and Astrophysics, Post Bag-4, Ganeshkhind, Pune 411007, India; girjesh@iucaa.in}
\affil{\altaffilmark{2}Physical Research Laboratory, Navrangpura, Ahmedabad 380009, India}
\email{{$^3$} Current address: DAMTP, University of Cambridge, UK; gg454@cam.ac.uk}

\begin{abstract}

Using the observations recorded by Atmospheric Imaging Assembly (AIA) on-board the Solar Dynamics Observatory (SDO), the Interface Region Imaging Spectrograph (IRIS) and the Extreme-ultraviolet Imaging Spectrometer (EIS) and X-Ray Telescope (XRT) both on-board Hinode, we present the evidence of chromospheric evaporation in a coronal loop after the occurrence of two active region transient brightenings (ARTBs) at the two footpoints. The chromospheric evaporation started nearly simultaneously in all the three hot channels of AIA such as 131~{\AA}, 94~{\AA} and 335~{\AA}, which was observed to be temperature dependent, being fastest in the highest temperature channel. The whole loop became fully brightened following the ARTBs after $\approx25$~s in 131~{\AA}, $\approx 40$~s in 94~{\AA}, and $\approx 6.5$~min in 335~{\AA}. The DEM measurements at the two footpoints (i.e., of two ARTBs) and the loop-top suggest that the plasma attained a maximum temperature of $\sim$10~MK at all these locations. The spectroscopic 
observations from IRIS revealed the presence of redshifted emission of $\sim$20~km~s$^{-1}$ in cooler lines like \ion{C}{2} and \ion{Si}{4} during the ARTBs that was co-temporal with the evaporation flow  at the footpoint of the loop. During the ARTBs, the line width of \ion{C}{2} and \ion{Si}{4} increased nearly by a factor of two during the peak emission. Moreover, enhancement in the line width preceded that in the Doppler shift which again preceded enhancement in the intensity. The observed results were qualitatively reproduced by 1-D hydrodynamic simulations where energy was deposited at both the footpoints of a monolithic coronal loop that mimicked the ARTBs identified in the observations.
\end{abstract}


\section{Introduction}
Active regions transient brightenings \citep[ARTB,][]{1992PASJ...44L.147S} are small intense events occurring in or near the active regions and were first studied using a sequence of images recorded by the Soft X-ray Telescope \citep[SXT,][]{1991SoPh..136...37T} on-board \textsl{Yohkoh}. These events are mainly found in active regions, on an average, every 3 min and have various morphologies like single-point or multiple-point brightening. ARTBs have also been observed with simultaneous multi-loop brightenings or even jets \citep{1994ApJ...422..906S,2014A&A...561A.104C,2016A&A...589A..79M}. Moreover, they have been observed as precursor to prominence eruptions \citep[e.g.,][]{2006A&A...453.1111T, 2006A&A...458..965C, 2007A&A...472..967C}, prominence oscillations \citep[e.g.,][]{2006A&A...449L..17I, 2009SSRv..149..283T} etc. They may play an important role in the heating of solar corona \citep{1995Natur.375...42Y}. Therefore, it is imperative to fully comprehend the thermodynamics involved in such 
events with the aid of modern high-resolution imaging and spectroscopic observations and modeling.

\citet{2001A&A...369..291B} investigated the correspondence between X-ray observations of ARTBs with EUV brightening. It was found that the strongest ARTBs observed in EUV had a counterpart in X-rays, whereas weaker EUV ARTBs did not show any X-ray emission. This was attributed to the fact that no plasma is heated beyond 2~MK. Since the energy values of ARTBs were similar to micro-flares, i.e., six orders of magnitude lower than a flare, ARTBs are also referred as micro-flares \citep[see,][]{hudson2004}. Hereafter, we shall be using micro-flares and ARTBs terms interchangeably.

Using the Hard X-ray observations recorded by The Reuven Ramaty High Energy Solar Spectroscopic Imager \citep[RHESSI,][]{2002SoPh..210....3L}, \citet{2008ApJ...677.1385C} performed a detailed statistical study of micro-flares and obtained frequency distributions of the peak energy \citep[see also][using Hi-C and AIA observations]{2014ApJ...784..134R, Sub_KT}. These distributions can be represented by power-law distributions with a negative power-law index between 1.5{--}1.6.  Such distributions suggest that the process of energy release in micro-flares is similar to those in large flares, though the energy involved is about six orders of magnitude less. The process of the energy release in such micro-flaring events or flaring events can be summarized as follows.

\begin{figure*}[htbp] 
\centering
\includegraphics[width=0.6\textwidth]{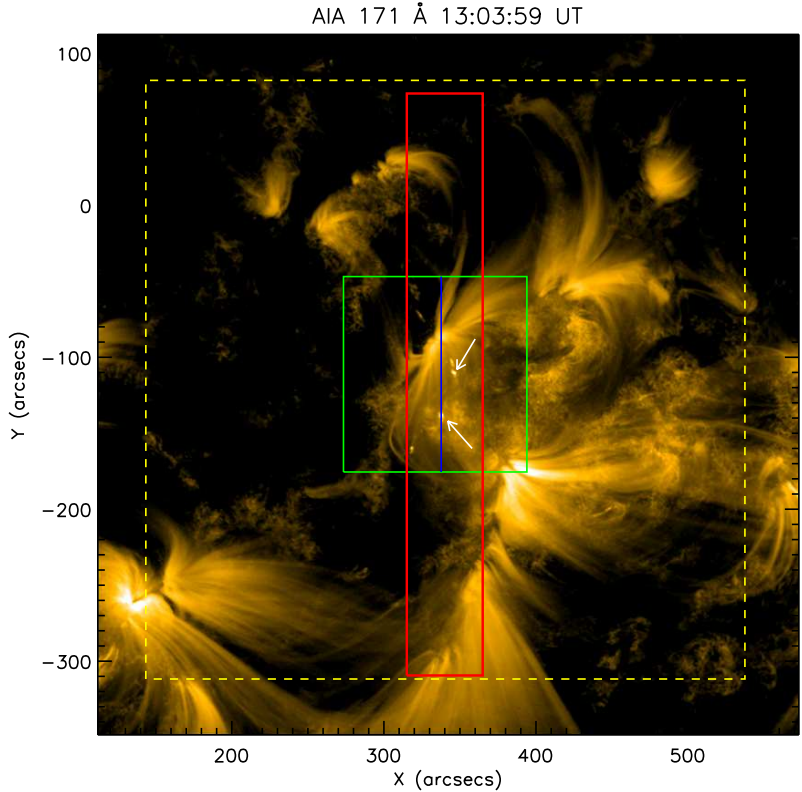}
\caption{AIA 171~{\AA} image recorded on 4 March 2014 shows the Active region transient brightenings (ARTBs) marked with arrows. The over-plotted red and green boxes show the EIS and IRIS-SJI field of views respectively. The yellow box shows the XRT FOV. The IRIS slit position is shown with the vertical blue line.}
\label{fig:obs}
\end{figure*}

At the time of impulsive energy release in (micro-)flares, local chromospheric material gets heated up to very large temperatures $\approx 10$ MK. This leads to increase in local pressure that drives the  local chromospheric material upward in to the corona along the (micro-)flare loops. Such process of filling of the loop with hot plasma is called \textquoteleft chromospheric evaporation\textquoteright\ \citep[][and references therein]{2009ApJ...699..968M,2017ApJ...841L...9L}. The concept of chromospheric evaporation have been successfully used to explain the loop-top brightening in post-flare loops \citep[see, e.g.][]{2016ApJ...823...47S} as well as coronal loops in general \citep[][]{2006SoPh..234...41K}. Along with this, high pressure also pushes denser plasma downwards into the lower chromosphere, which is referred as \textquoteleft chromospheric condensation\textquoteright\ \citep[][and references therein]{2006ApJ...638L.117M}. In such models, emissions from hot materials from corona are expected to 
show blueshift (upflow) and that from cool material from upper chromosphere and transition region are expected to show redshift (downflow) as reported recently in a single observation by \citet{2017ApJ...841L...9L}. As underlying chromosphere is denser than the overlying corona, the expected speed of blueshift (upflow) should be much larger (approximately an order of magnitude) than  that of redshift (downflow) \citep[e.g.,][]{1985ApJ...289..414F,2009ApJ...699..968M}. The spectroscopic observations do show larger blueshifts for the spectral lines formed at higher temperatures \citep{2006SoPh..234...95D}. Such effects have also been seen at the footpoints of loops in quiescent active regions and are interpreted as evidence of nanoflares occurring in the corona \citep[see e. g.,][ and references therein]{2009ApJ...694.1256T, 2017ApJ...835..244G}.

\begin{figure*}[hbtp]
\centering
\includegraphics[width=0.98\textwidth]{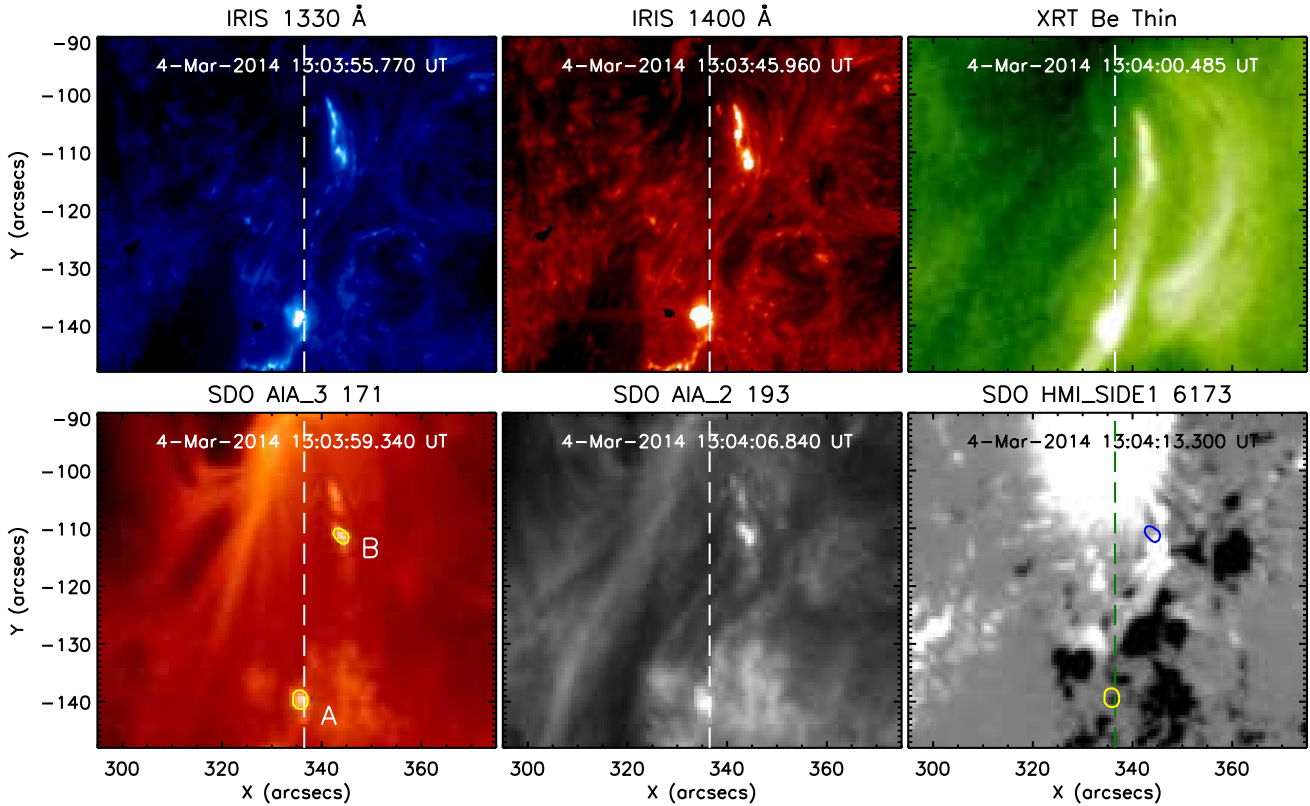}
\caption{Images obtained from IRIS slit-jaw, XRT, and AIA, showing the active region transient brightenings (ARTBs). The different passbands are labeled. Corresponding HMI magnetic field map scaled between $\pm$ 80~G is shown in the bottom right panel. The two contours on HMI image indicate brightening location obtained from AIA 171~{\AA} image. The vertical dashed lines on all images represent the location of IRIS slit.}
\label{fig:context_loc}
\end{figure*}

\begin{figure*}[htbp] 
\centering
\includegraphics[width=0.95\textwidth]{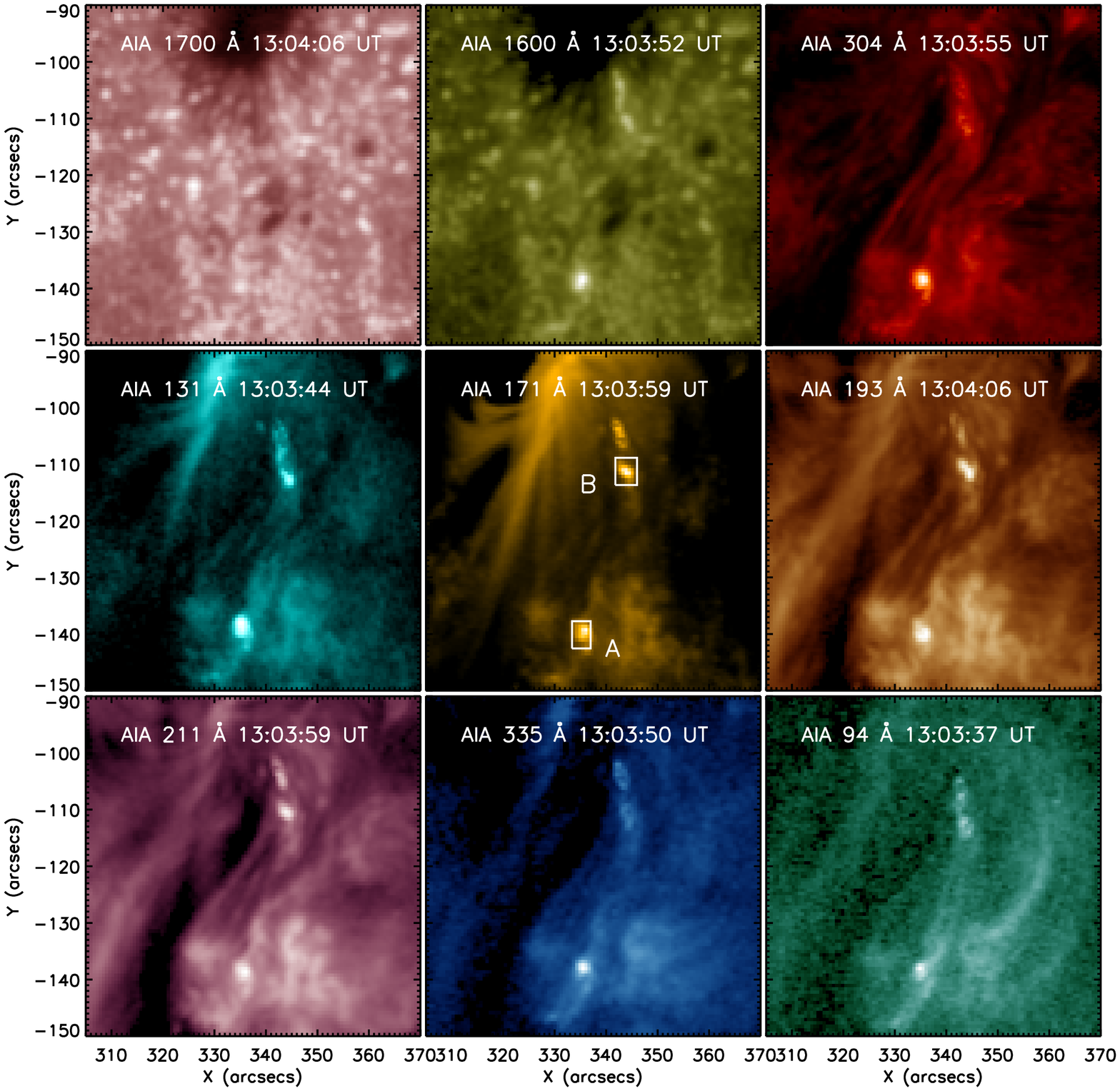}
\caption{AIA images show the active region transient brightenings (ARTB) at the peak of its life. The two boxes on AIA~171~{\AA} locate the ARTBs as identified in AIA 171~{\AA} image.}
\label{fig:aia_img}
\end{figure*}

As solar corona has high thermal conductivity and low plasma-$\beta$, plasma is confined within the magnetic field lines. Thus, the evolution of plasma can be well demonstrated with 1-D hydrodynamic model \citep[see, e.g.,][]{1987ApJ...319..465M, 2000SoPh..193...33P, 2008ApJ...683..516S}. In the current study, we present observations of two ARTBs and associated loop brightening by Atmospheric Imaging Assembly \citep[AIA;][]{2012SoPh..275...17L}, Interface Region Imaging Spectrograph \citep[IRIS;][]{2014SoPh..289.2733D} and Extreme-ultraviolet Imaging Spectrometer \citep[EIS;][]{2007SoPh..243...19C} 
and X-Ray Telescope \citep[XRT;][]{2007SoPh..243...63G} both on-board Hinode. Based on this observation, we set up a 1-D hydrodynamic simulation to study the dynamic evolution of the coronal loop using the procedure described by \cite{2008ApJ...683..516S}.

The rest of the paper is organized as follows. In Section~\ref{sec:obs}, we present the observations of ARTBs and the associated loop. Physical properties of the event is described in Section~\ref{sec:analysis}. The hydrodynamic simulation of loop and its comparison with observational results are described in Section~\ref{sec:model}. Finally, we summarize and discuss our results in Section~\ref{sec:summary}.

\section{Observations} \label{sec:obs}

The ARTB event under study occurred on 2014 March 4 at around 13:03 UT. The event was simultaneously recorded using multiple instruments such as the AIA and the Helioseismic and Magnetic Imager \citep[HMI;][]{2012SoPh..275..207S,2012SoPh..275..229S} both on-board the Solar Dynamics Observatory (SDO), EIS \& XRT on-board Hinode, and IRIS. IRIS was operating in the sit-and-stare mode while EIS was rastering the region of interest. Figure~\ref{fig:obs} displays the AIA image taken in 171~{\AA}. Over-plotted green and red boxes highlight the EIS raster and IRIS/SJI field of view (FOV) respectively. XRT field of view is indicated by an over-plotted dashed yellow box. The blue vertical line represents the IRIS slit position. The arrows mark the ARTBs, which are the subjects of this work. 

In the current study, IRIS observations were carried out from 12:39:30 UT to 14:37:54 UT and covered the field of view of 119$\arcsec \times$ 119$\arcsec$
EIS raster was obtained between 13:02:20 UT and 13:20:23 UT and  covered the field of view of 51$\arcsec \times$ 384$\arcsec$ using the 1$\arcsec$ slit.
IRIS obtained spectra in the wavelength windows of lines \ion{C}{2} 1335 \& 1336~{\AA}, \ion{O}{1} 1355~{\AA}, \ion{Si}{4} 1394 \& 1403~{\AA} whereas
EIS had good signal strength only for spectral lines \ion{Fe}{12}~186 and 195~{\AA}, and \ion{Fe}{15} 284~{\AA} among the observed lines.
The exposure time for IRIS in spectroscopic mode was 4~s that resulted into an effective cadence of 5~s in sit-and-stare mode. The IRIS-SJI were obtained with
an exposure time of 4~s and effective cadence of 18~s, 15~s, and 15~s for SJI 1330~{\AA}, 1400~{\AA}, and 2796~{\AA}, respectively.
EIS spectra were obtained with an exposure time of 20~s. XRT images were obtained in several filters such as Al\_poly, C\_poly, Be\_thin, Be\_med, Al\_med, Ti\_poly, and Al\_thick  with varying exposure time and hence the cadence. IRIS and AIA observations were co-aligned using IRIS-SJI 1400~{\AA}, and AIA 1600~{\AA} images. All the HMI and AIA images obtained in different passbands were co-aligned and de-rotated to 
AIA 1600~{\AA} image obtained at 12:53:28~UT using the standard procedures provided in the Solar Software package \citep[SSW;][]{1998SoPh..182..497F}.
 We have used IRIS level-2 data where slit-jaw images from different filters and detectors are already co-aligned. We used IRIS cool neutral line \ion{S}{1} 1401.514~{\AA} to perform the absolute wavelength calibration of IRIS spectral lines. We followed standard procedures for calibrating the EIS data  using IDL routine EIS\_PREP available in the SSW. We fitted the EIS spectral line profiles with Gaussian function using EIS\_AUTO\_FIT. Spatial offset between EIS images obtained from different wavelengths  were corrected with respect to image obtained from \ion{Fe}{12}~195.12 {\AA} spectral line.
 
\begin{figure*}[htbp]  
\centering
\includegraphics[width=0.8\textwidth]{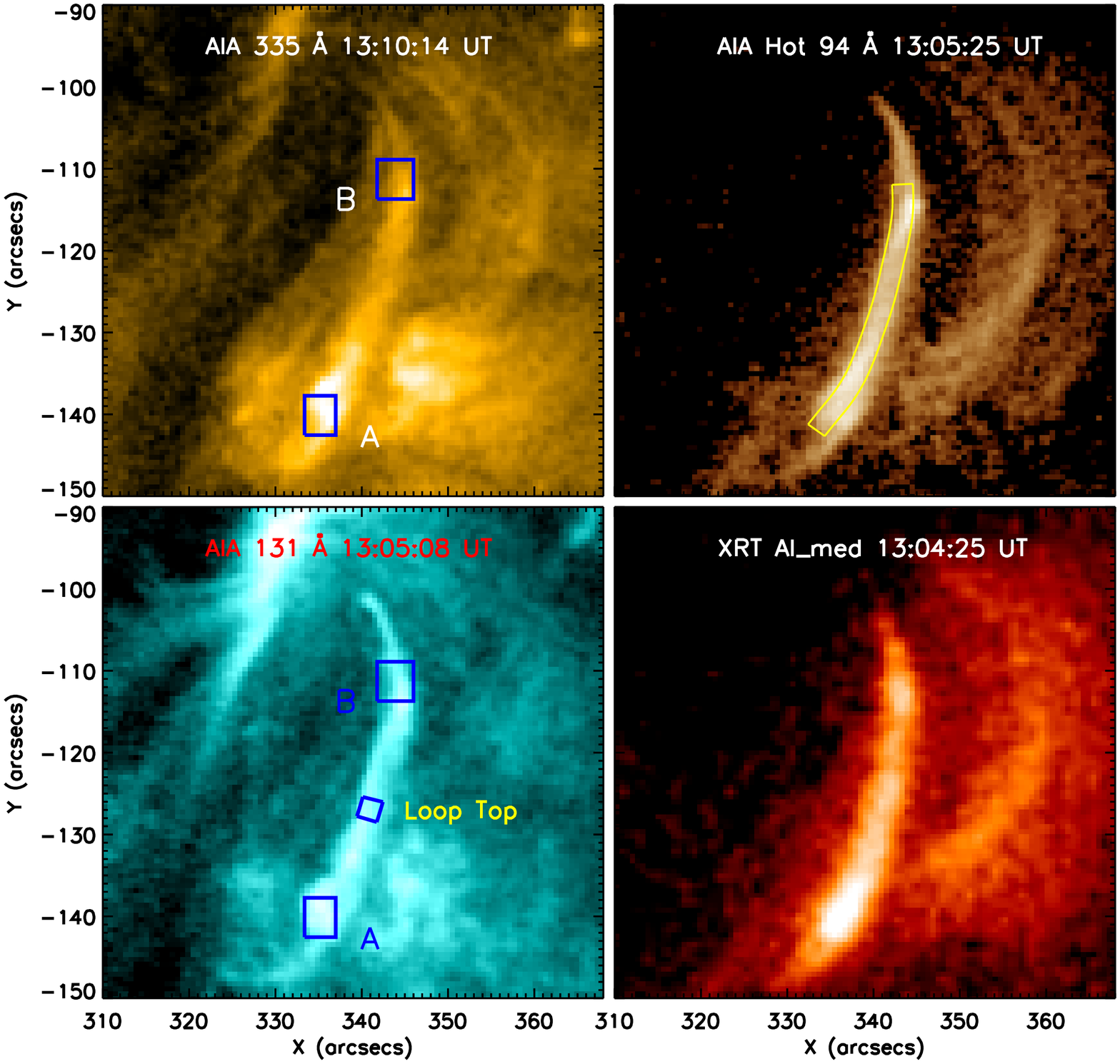}
\caption{The brightened loop after the ARTB at the peak time of its evolution showed in AIA 335~{\AA} (top left), corrected 94~{\AA} (top right), 131~{\AA} (bottom left) and in XRT Al med filter (bottom right). Corrected AIA 94~{\AA} image is obtained by subtracting the contributions from cool temperature components (see the text for details). The over-plotted 5-pixel wide box in top right panel traces the hot loop. The two blue boxes \textquoteleft A\textquoteright\ and \textquoteleft B\textquoteright\ locate the ARTBs, i.e., foot-points of brightened loop whereas loop-top is also marked with the blue box as labeled.}
\label{fig:hot_loop}
\end{figure*}

Figure~\ref{fig:context_loc} shows the locations of the ARTBs in IRIS-SJIs recorded at 1330~{\AA} and 1400~{\AA}, XRT Be-thin filter image, AIA 171~{\AA} and 193~{\AA} images. Corresponding line-of-sight (LOS) magnetic field map obtained from HMI is plotted in the bottom right panel. \textquoteleft A\textquoteright\ and \textquoteleft B\textquoteright\ in the bottom left panel locates the two ARTBs that appear simultaneously. Both the ARTBs appeared and peaked at the same time in different passbands of IRIS and AIA. Intensity contours corresponding to the ARTBs obtained from AIA 171~{\AA} at its peak are over-plotted on the HMI LOS magnetic map. The region covered by these contours show the presence of a strong gradient in the magnetic field underneath ARTBs as compared to its surroundings. At point \textquoteleft A\textquoteright, the average strength of LOS magnetic field is about $-400$~G, which is surrounded by an average magnetic field strength of $-50$~G. Point \textquoteleft B\textquoteright , 
shows strong gradient in the field strength where one half shows field strength of 650 G whereas another half of 120 G. As estimates are made over only LOS magnetic field, such sharp gradient may indicate that these regions are possibly located in the opposite polarity regions. The vertical dashed lines represent the IRIS spectroscopic slit that passes above the ARTB \textquoteleft A\textquoteright . 

\section{Data Analysis and Results} \label{sec:analysis}
\subsection{Imaging Analysis}
\label{sec:imaging}

\begin{figure*}[htbp] 
\centering
\includegraphics[width=0.9\textwidth]{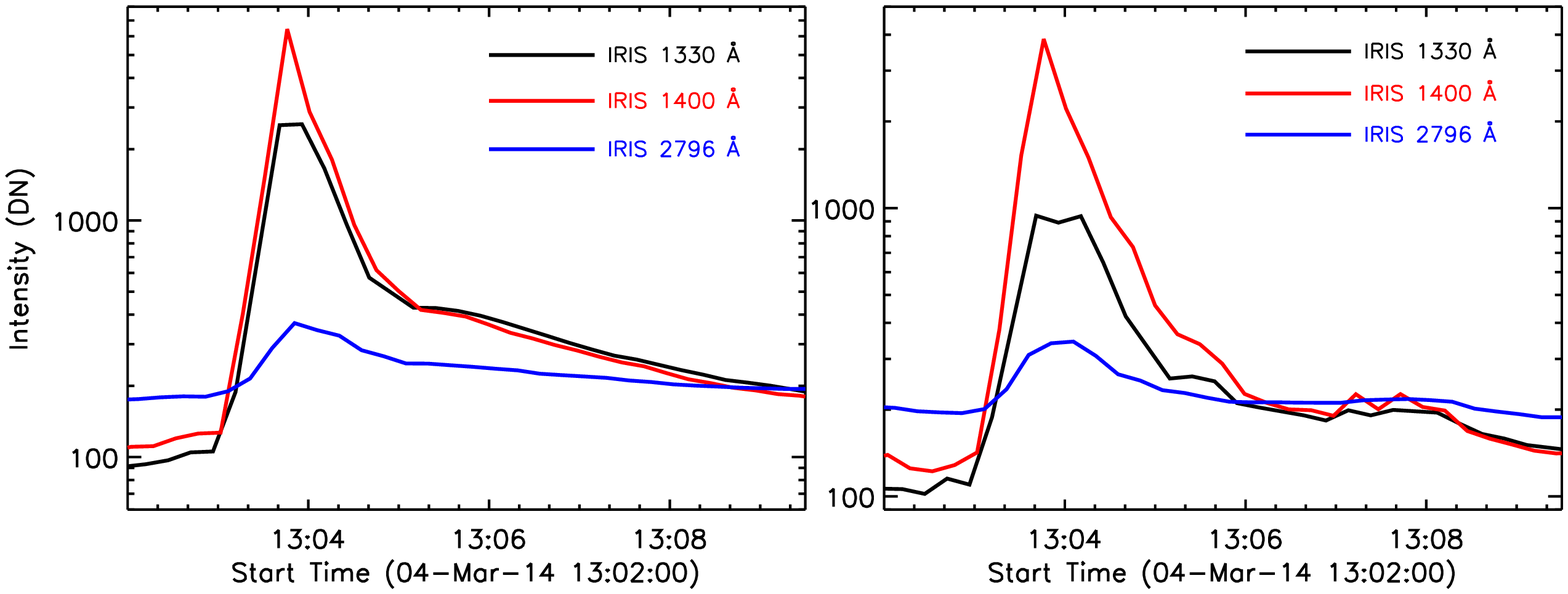}
\caption{Light curves obtained using IRIS-SJI at ARTBs \textquoteleft A\textquoteright\ (left panel) and \textquoteleft B\textquoteright\ (right panel).} \label{fig:lc_iris}
\end{figure*}

\begin{figure*}[htbp] 
\centering
\includegraphics[width=0.8\textwidth]{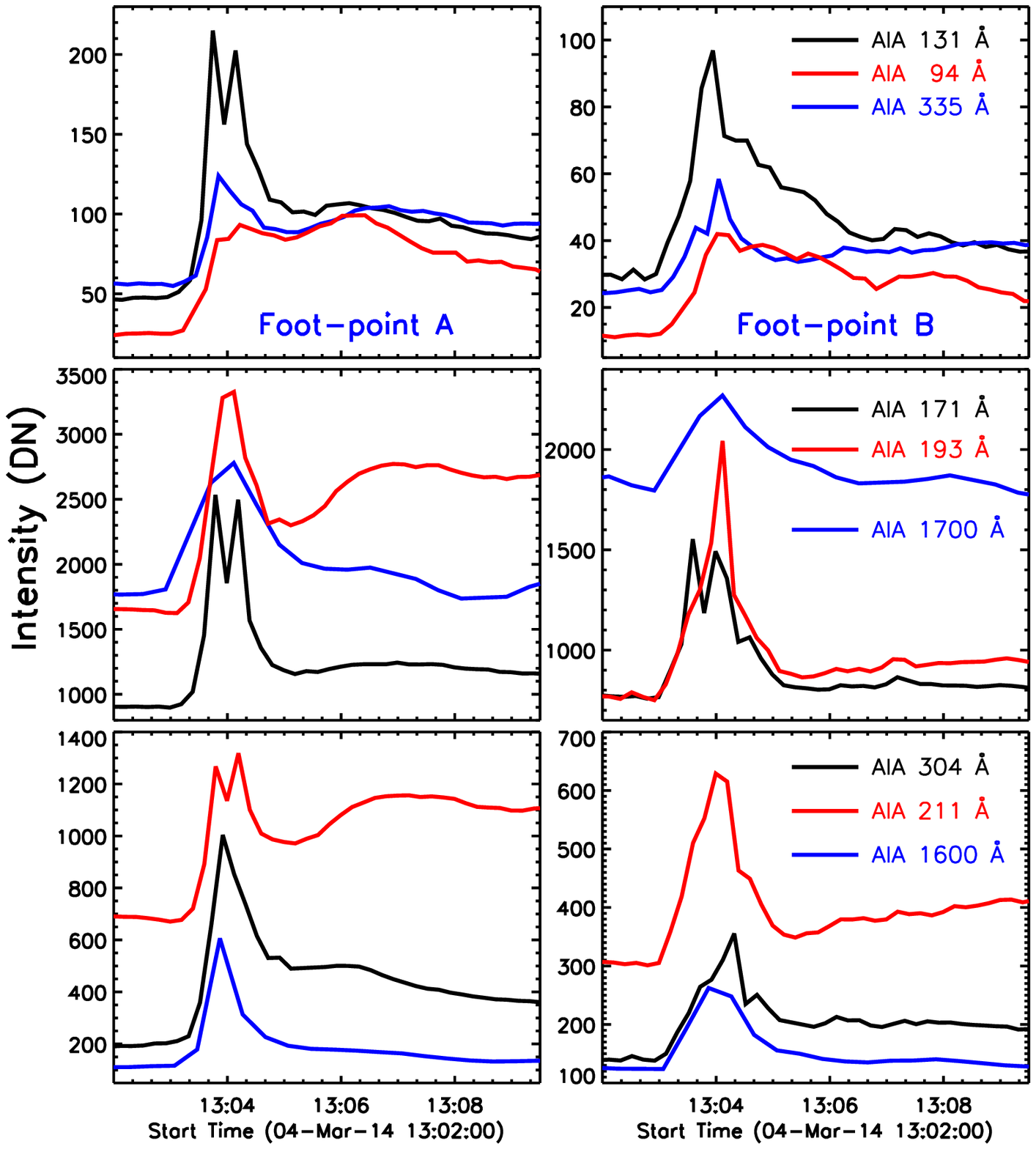}
\caption{Light curves obtained using all the AIA passbands corresponding to the ARTBs \textquoteleft A\textquoteright\ (left panels ) and  \textquoteleft B\textquoteright\ (right panels).}
\label{fig:lc_aia}
\end{figure*}

The two ARTBs namely \textquoteleft A\textquoteright\ and \textquoteleft B\textquoteright\, having compact sizes of about $2\arcsec \times 2\arcsec$\, started at $\approx$13:03:11 UT and peaked at around 13:03:45~UT. Both \textquoteleft A\textquoteright\ and \textquoteleft B\textquoteright\ were recorded in all the AIA and IRIS-SJI passbands. Figure~\ref{fig:aia_img} displays the two ARTBs at their peak recorded in AIA passbands. The ARTBs are also observed by XRT using Be-thin filter as shown in the top right panel of Figure~\ref{fig:context_loc}. The two channels of AIA namely 131~{\AA}, and 94~{\AA} are sensitive to hot plasma representing \ion{Fe}{21} 128.75~{\AA} and \ion{Fe}{18} 93.93~{\AA} respectively. However, both of these channels have strong contributions from cooler temperatures \citep{2010A&A...521A..21O, 2011A&A...535A..46D}. Fortunately, emission due to hot plasma corresponding to \ion{Fe}{18}~93.93~{\AA}  spectral line from 94~{\AA} channel images can be separated from cool emission \citep{
2012ApJ...759..141W, 2013A&A...558A..73D}. In the current study, we have utilized the approach of \citet{2013A&A...558A..73D} where the emission due to the hot \ion{Fe}{18} line can be obtained as:

\begin{equation}
 I(Fe~XVIII) \approx I(94~{\AA}) - I(211~{\AA})/120 - I(171~{\AA})/450
\end{equation}

where  I(94~{\AA}), I(211~{\AA}), and I(171~{\AA}) are intensities from AIA 94~{\AA}, 211~{\AA}, and 171~{\AA} passbands respectively. 

After the first appearance of ARTBs, a loop structure was observed at different times in different filters with the footpoints located at the two ARTBs namely \textquoteleft A\textquoteright\ and \textquoteleft B\textquoteright. Figure~\ref{fig:hot_loop} displays the hot loops seen in three AIA channels namely AIA 335~{\AA}, corrected 94~{\AA} (\ion{Fe}{18}), and AIA 131~{\AA} and in XRT Al\_med filter. The full extent of the hot loop was observed in AIA 131, corrected 94, and 335~{\AA} filters at times 13:05:08, 13:05:25, and 13:10:14~UT respectively, whereas the peak phase of ARTB appeared at time 13:03:44~UT as observed from AIA 131~{\AA} passband. We have identified and traced the complete hot loop in the corrected AIA 94~{\AA}  (top right panel in Figure~\ref{fig:hot_loop}) passband from foot-points \textquoteleft A\textquoteright\ to \textquoteleft B\textquoteright\  for further analysis. The traced loop is over-plotted in the top right panel of Figure~\ref{fig:hot_loop}. The width of the  traced loop 
is chosen to be 5-pixels of AIA.

\begin{figure*}[htbp] 
\centering
\includegraphics[width=0.9\textwidth]{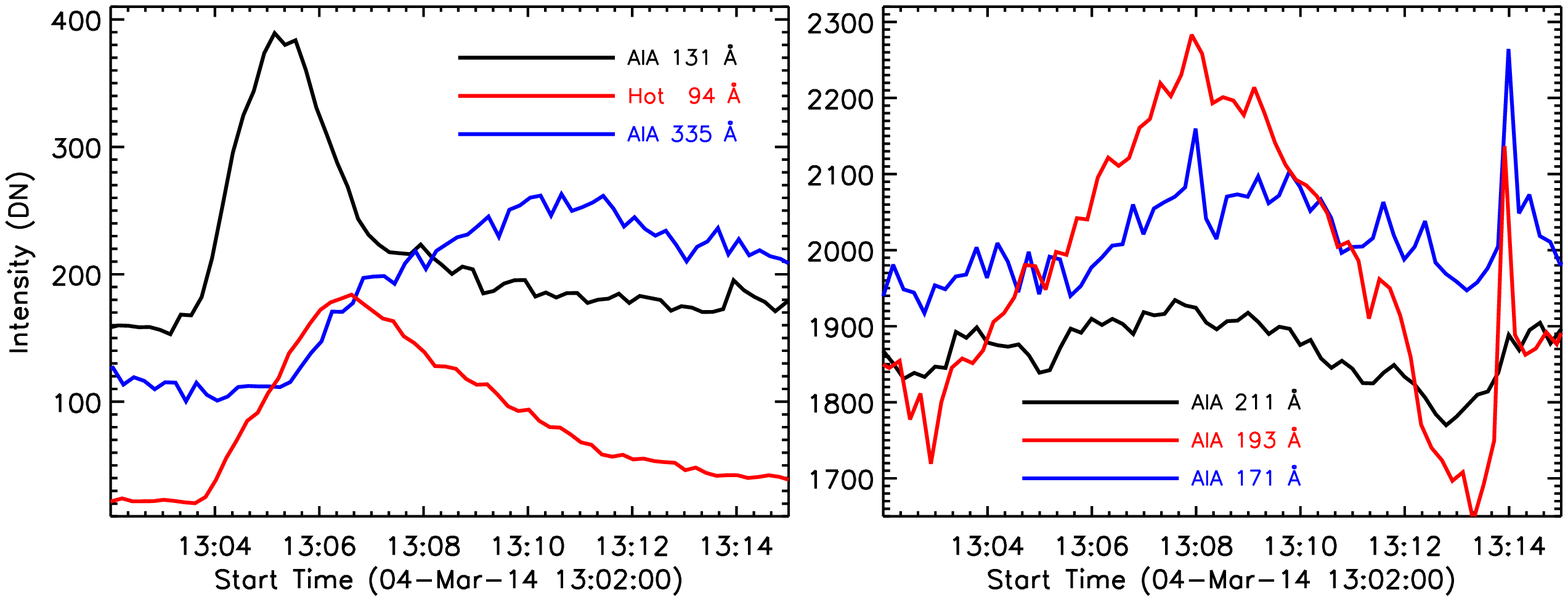}
\caption{Light curves obtained at the loop-top using all the AIA passbands.}
\label{fig:lc_top}
\end{figure*}

\begin{figure*}[htbp] 
\centering
\includegraphics[width=0.9\textwidth]{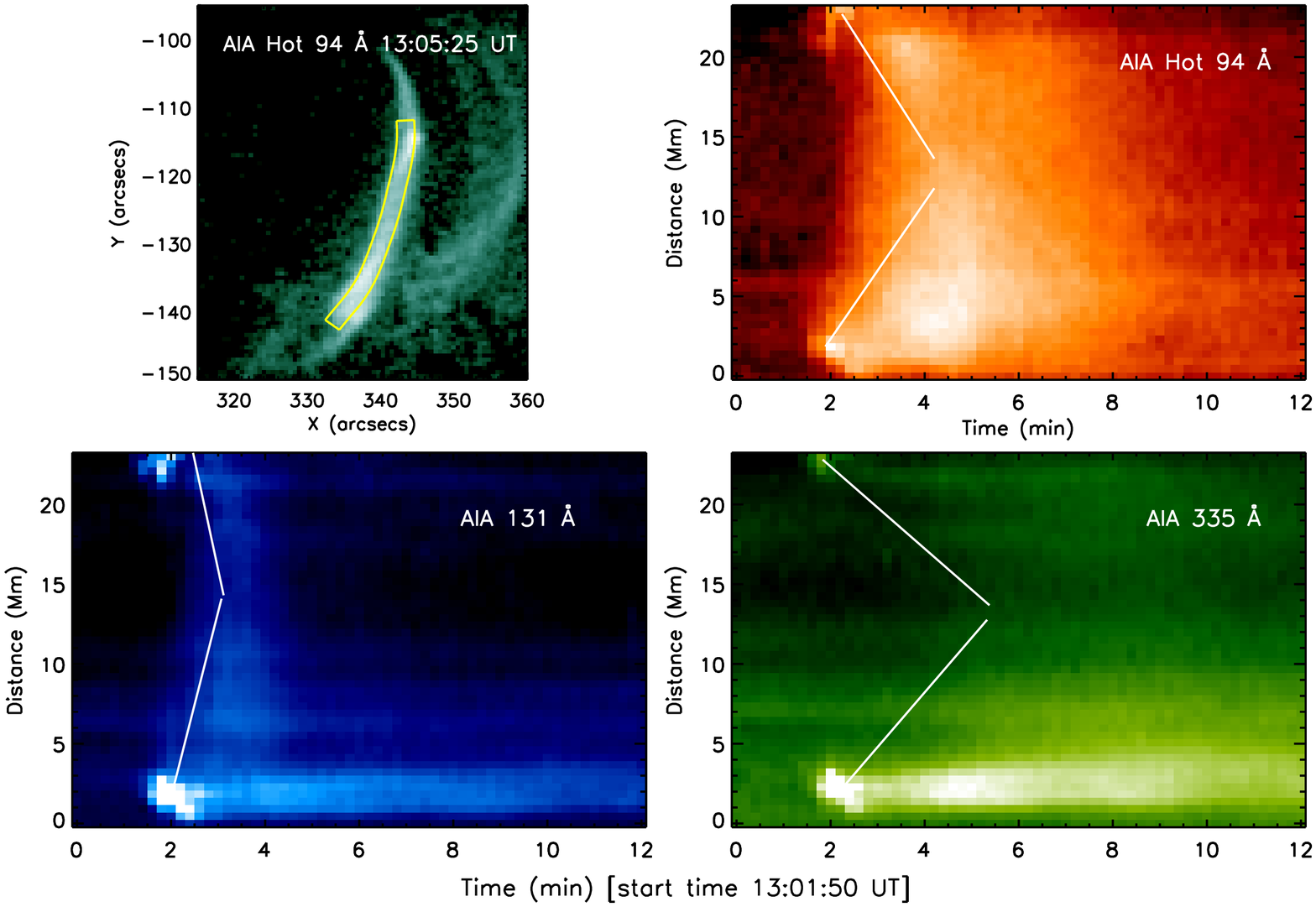}
\caption{Top left panel: The brightened loop as observed in corrected AIA 94~{\AA} over-plotted with a 5-pixel box. The other panels show the time-distance plots obtained from corrected 94~{\AA}, 131~{\AA},
and 335~{\AA}. The two white lines are drawn to guide the eyes to flows. Hot plasma flows from both foot-points \textquoteleft A\textquoteright\ and \textquoteleft B\textquoteright\ towards the loop-top.} 
\label{fig:bright_front}
\end{figure*}

Figure~\ref{fig:lc_iris} plots the light curves of ARTBs \textquoteleft A\textquoteright\ (left panel) and \textquoteleft B\textquoteright\ (right panel) obtained from all the three IRIS-SJIs taken in 1330~{\AA}, 1400~{\AA} and 2796~{\AA}. For the ARTB \textquoteleft A\textquoteright\ (\textquoteleft B\textquoteright ), the light curves show enhancements from the initial value in the intensities by a factor of more than 25(30), 60(8), and 2(1.7) in 1330~{\AA}, 1400~{\AA}, and 2796~{\AA} respectively. It is noted that for \textquoteleft A\textquoteright\, the intensity in 1330~{\AA} filter becomes saturated during its peak. Both the ARTBs attained their peak within the 45~s (i.e., within the three time-frames) of its start and became visible nearly simultaneously in all the IRIS-SJIs (within the error of different observation times for different filters). 

The light curves obtained for \textquoteleft A\textquoteright\ and \textquoteleft B\textquoteright\ using all the AIA channels is plotted in Figure~\ref{fig:lc_aia}. As can be inferred from the plots, the intensities for both the ARTBs are enhanced by about a factor of 2 to 5 in different AIA passbands from their initial values. Both ARTBs attained their peak intensities within 36~s (i.e., within the three time-frames) of its start. Similar to IRIS, the ARTBs attained their peak nearly simultaneously in all the channels (within the error of different observation times for different filters). We further note here the double-peaked light curves in some channels.

To better understand the correlation between the ARTBs (shown in Figure~\ref{fig:aia_img}) and the appearance of the hot loops in AIA channels (shown in Figure~\ref{fig:hot_loop}), we plot the light curves obtained at the loop-top for hot channels like 131~{\AA}, corrected 94~{\AA}, and 335~{\AA} in the left panel of Figure~\ref{fig:lc_top}. The loop-top light curves for cooler channels are plotted in the right panel of Figure~\ref{fig:lc_top}. Position of the loop-top is shown in Figure~\ref{fig:hot_loop}. As can be inferred from these light curves,  intensities at the loop-top are enhanced by the factors of about 1.4, 2.2, and 9.0 in AIA 335~{\AA}, 131~{\AA}, and 94~{\AA} channel corrected for \ion{Fe}{18} emissions, respectively. Unlike the simultaneous appearance of the peak of ARTBs in all the channels of IRIS and AIA,  the loop-top intensities in different channels attain their peak at different times. The left panel reveals that the loop-top intensity first peaks in 131~{\AA} channel and followed in 
corrected 94~{\AA}, and then 335~{\AA}. The light curves for the cooler channels do not show strong variation in the loop-top intensities.

Using the traced loop shown in Figure~\ref{fig:hot_loop}, we created time-distance map using corrected 94~{\AA} (top right), AIA 131~{\AA} (bottom left) and 335~{\AA} (bottom right) channels, and plotted in Figure~\ref{fig:bright_front}. These time-distance plots clearly reveal the movement of hot plasma from footpoint regions to loop-top. The white over-plotted lines on time-distance plots are drawn to visualize the plasma movement. The steepness of the white lines in different time-distance plots, suggests that the hot loop is first seen in 131~{\AA} channel followed by corrected 94~{\AA} channel, and later in  335~{\AA} channel. This is exactly observed in the light curves shown in Figure~\ref{fig:lc_top}. The whole loop became fully brightened after the ARTBs after $\approx25$~s in 131~{\AA}, $\approx 40$~s in 94~{\AA}, and $\approx6.5$~min in 335~{\AA}. We estimated the speed of plasma motion from foot-points towards the lop top by measuring the slope of these white-lines. We found that the speed was 
highest in 131~{\AA} ($\approx 211$~km~s$^{-1}$), followed by corrected 94~{\AA} ($\approx 74$ ~km~s$^{-1}$, and 335~{\AA} ($\approx 50$  km~s$^{-1}$). We note that such an analysis from XRT observations was not possible as continuous observations for individual XRT filters were not available.

\begin{figure}[htbp] 
\centering
\includegraphics[width=0.5\textwidth]{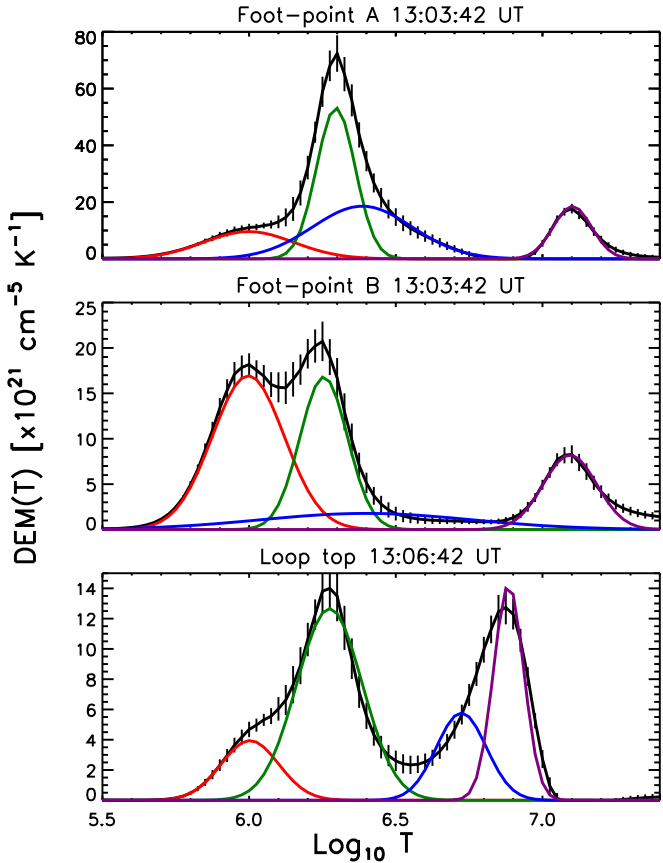}
\caption{DEM curves obtained using six AIA coronal passbands at  ARTBs \textquoteleft A\textquoteright\ (top panel), \textquoteleft B \textquoteright\ (middle panel), and loop-top (bottom panel). 
DEM profiles are fitted with 4 Gaussian functions and polynomial of degree 2.}
\label{fig:aia_dem}
\end{figure}

We further study the evolution of temperature at the locations of ARTBs as well as along the hot loop that is rooted at the two ARTBs. For this purpose, we performed a DEM analysis by employing the method of \citet{2012A&A...539A.146H} on the six coronal channels of AIA namely 94, 131,171, 193, 211, and 335~{\AA}. Figure~\ref{fig:aia_dem} displays the DEM curves obtained at ARTB \textquoteleft A\textquoteright\ (top panel) and \textquoteleft B\textquoteright\ (middle panel), and at the loop-top (bottom panel) during their peak emission. The DEM curves obtained at both foot-points \textquoteleft A\textquoteright\  and \textquoteleft B\textquoteright\  show similar features with respect to temperature. The DEM curves also suggest that both the ARTBs as well as the loop-top show multi-thermal emission.
 
\begin{figure}[htbp] 
\centering
\includegraphics[width=0.5\textwidth]{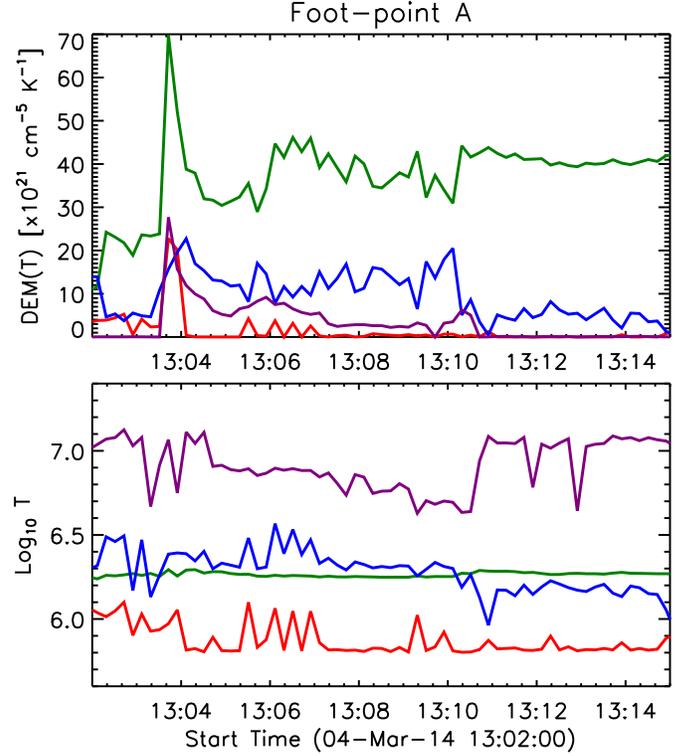}
\caption{Time evolution of DEM components (top panel) and corresponding temperature (bottom panel) at ARTB \textquoteleft A\textquoteright . The Lines are color-coded as same to that of 4 Gaussian components
used to fit DEM curves as shown in the top panel of Figure~\ref{fig:aia_dem}.}
\label{fig:temp_A}
\end{figure}

To separate out hot plasma components from the background at foot-points and along the loop, we fitted the DEM curves with 4 Gaussian functions and polynomial of degree 2, similar to the approach employed by \citet{2011ApJ...732...81A}. Chosen function best represents the DEM curves for the desired purpose. These fitted profiles also provide the temperature of different plasma components. The four Gaussian components represent cool ($\approx 10^{5.8}-10^6$ K), warm ($\approx 10^{6.2}-10^{6.3}$ K), intermediate ($\approx 10^{6.2}-10^{6.7}$ K), and hot ($\approx 10^{6.7}-10^{7.2}$ K) temperature plasma components. These Gaussian functions representing different temperature components are over-plotted with different colors in Figure~\ref{fig:aia_dem}. We use red and green color to represent the cool and warm background plasma, respectively. The dark blue color represents newly emerged hot plasmas due to the ARTBs. The intermediate temperature plasma component was added to obtain the best fit for the DEM curves 
and are represented with blue color. Although this component is added only to full fill the best fit requirement, it can be interpreted as the appearance of intermediate hot plasma component. This plasma component initially appears with hot plasma component and later merges with warm background plasma component.

\begin{figure}[htbp] 
\centering
\includegraphics[width=0.48\textwidth]{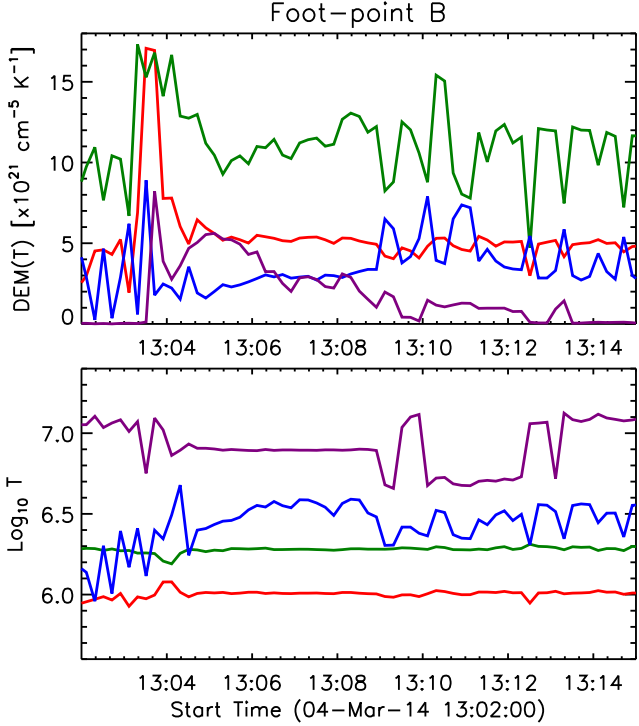}
\caption{Same as Figure~\ref{fig:temp_A} but for ARTB \textquoteleft B\textquoteright .}
\label{fig:temp_B}
\end{figure}

\begin{figure}[htbp] 
\centering
\includegraphics[width=0.48\textwidth]{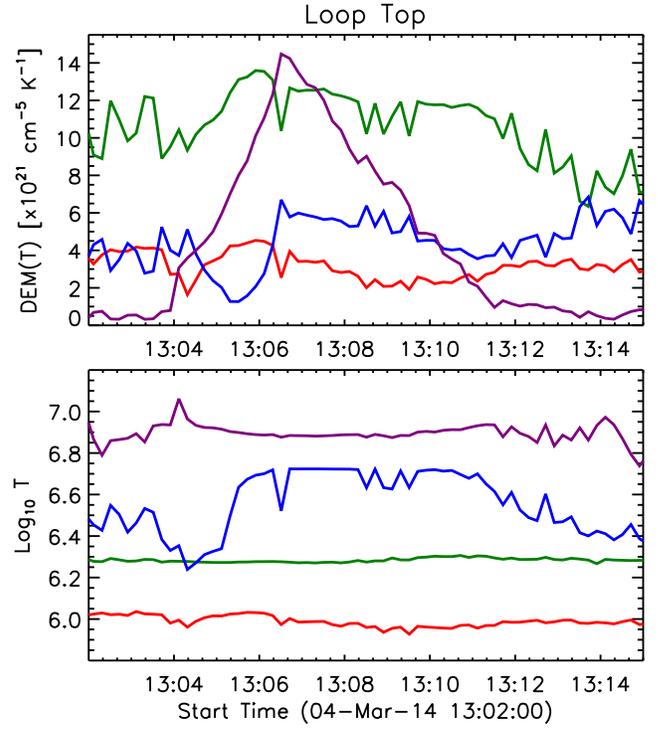}
\caption{Same as Figure~\ref{fig:temp_A} but for the loop-top.}
\label{fig:temp_top}
\end{figure}

Figures~\ref{fig:temp_A}, \ref{fig:temp_B}, and ~\ref{fig:temp_top} show the time evolution of peak values of DEMs obtained at different temperature components (top panels) and associated temperatures (bottom panels) obtained at ARTBs \textquoteleft A\textquoteright , \textquoteleft B\textquoteright, and at the loop-top, respectively. The color of the curves shown in Figures~\ref{fig:temp_A}, \ref{fig:temp_B} and ~\ref{fig:temp_top} are same as those for Gaussians shown in Figure~\ref{fig:aia_dem}. These curves again reveal the existence of background cool and warm plasma components throughout the time evolution of the ARTB \textquoteleft A\textquoteright\ and  \textquoteleft B\textquoteright\ as well as the loop-top. The emission measure from warm plasma component is higher than that from cool plasma throughout the evolution. The contribution from hot plasma component due to temperature $\approx 10^{6.7}-10^{7.2}$~K appears only during the ARTB and later in the associated loop. The temperature of hot 
component of the plasma changes with time. In the beginning, at the ARTBs, the plasma heats up to a temperature of $\approx 10^{7.11}$~K at 13:04:07 UT and later drops continuously afterward and reaches a temperature of $\approx 10^{6.64}$ K after about 6 minutes. The plasma at the loop-top heats up to the temperature of $\approx 10^{7.06}$~K at time 13:04:07 UT and decreases to $\approx 10^{6.87}$~K within 5 minutes. However, the loop-top shows highest emission measure at 13:06:31~UT due to this hot temperature component. Intermediate temperature plasma component which was added to obtain the best fit did not provide any concluding information. In this method, as noted by \citet{2013ApJ...778..139S}, when the difference between two temperatures T$_1$-T$_2$ becomes similar to the sum of two half widths at half maximum i.e., 1.31($\sigma_1+\sigma_2$), isolation of two temperature components become difficult.  In our case, it is difficult 
to isolate different temperature components for the foot-points \textquoteleft A\textquoteright ,  \textquoteleft B\textquoteright , and loop-top, beyond 13:10, 13:08, and 13:10 UT, respectively.

We also obtain the density at  foot-point \textquoteleft A\textquoteright\ using the differential emission measures (DEM) during the peak phase of the transient . We integrate the DEMs over the temperature range of 10-15 MK (see top panel of Figure~\ref{fig:aia_dem}) and obtained emission measure (EM).  We obtained the electron number density to be $9\times10^{10}$ cm$^{-3}$ at \textquoteleft A\textquoteright\ by assuming a filling factor of unity and column depth of $2\arcsec$ (which is the extent of brightening).

\subsection{Spectroscopic Analysis}
\label{sec:spectroscopy}

\begin{figure*}[htbp] 
\centering
\includegraphics[width=0.85\textwidth]{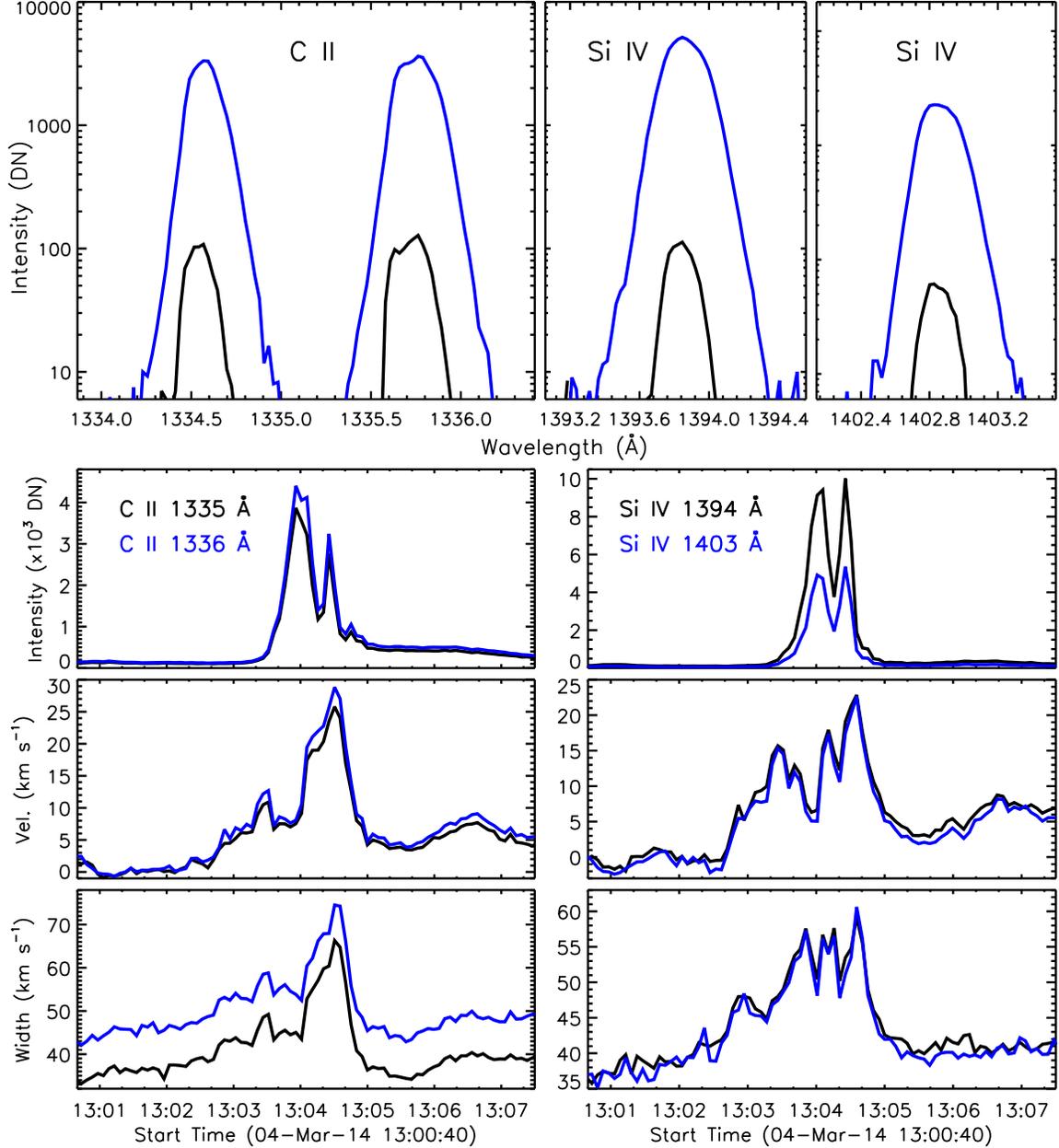}
\caption{Top panels: IRIS spectral line profiles obtained from \ion{C}{2}~1335 and 1336~{\AA}, and \ion{Si}{4}~1394 and 1403~{\AA} lines before (black line) and during the ARTB \textquoteleft A\textquoteright\ (blue line).
Time evolution of intensity, Doppler velocity, and line width obtained at \textquoteleft A\textquoteright\ using both the spectral lines are also plotted as labeled.}
\label{fig:prof}
\end{figure*}

Fortuitously, the spectroscopic slit during the sit-and-stare observation of IRIS was located right above the ARTB {\textquoteleft}A{\textquoteright} that allowed us to perform a detail spectroscopic study. The top row of Figure~\ref{fig:prof} displays the line profiles before (black) and during (blue) the event in \ion{C}{2}~1335 \& 1336~{\AA} and \ion{Si}{4}~1394 \& 1403~{\AA}. We fitted all the line profiles with a single Gaussian for further analysis. Figure~\ref{fig:prof} plots the extracted parameters from fitted Gaussians such as intensity (second row), Doppler velocity (third row), and line width (fourth row). Doppler velocity is estimated with respect to the
pre-event conditions. Intensity curves (second row) for  both \ion{C}{2} and \ion{Si}{4} (both the lines) show intensity enhancements by the factor of more than 35, and 100, respectively, during the event. We also note that several pixels were saturated at time 13:03:56~UT in both the spectral lines indicating the epoch of peak intensity, which 
was attained within 40~s of its start time as was also inferred from the IRIS-SJIs taken in 1330~{\AA} and 1400~{\AA}.

\begin{figure*}[htbp] 
\centering
\includegraphics[width=0.85\textwidth]{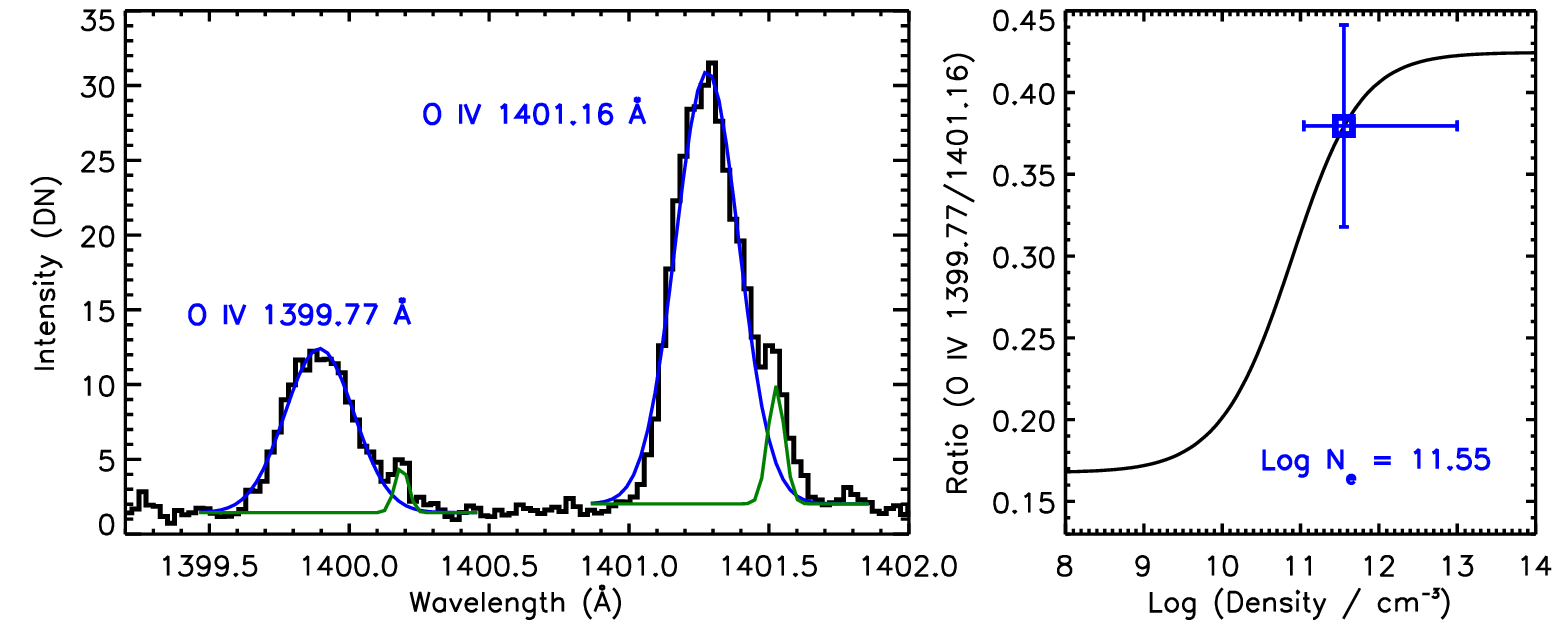}
\caption{Average electron number density estimated at ARTB  {\textquoteleft}A{\textquoteright} using IRIS  \ion{O}{4}~$\lambda$1399/$\lambda$1401 line pair during the transient event.}
\label{fig:density}
\end{figure*}

During the event, both the spectral lines of \ion{C}{2} and \ion{Si}{4} show strong redshifts i.e., downflows of $>25$ km~s$^{-1}$ and $>20$ km~s$^{-1}$, respectively. The spectral line widths also increased from $\approx32(40)$ km~s$^{-1}$ in the pre-event condition to more than $65(74)$ km~s$^{-1}$ during the event for \ion{C}{2}~1335(1336)~{\AA}. The difference of 8 km~s$^{-1}$ in the width of two lines of \ion{C}{2} could be attributed to the presence of several absorption features present within the wavelength span \citep{2015ApJ...809...82G}. The line width of two \ion{Si}{4} lines increased from $\approx35$ km~s$^{-1}$ to about 60 km~s$^{-1}$. A close  examination of time of enhancement in intensities, Doppler shifts, and line widths reveal that the line widths started to increase about 2 to 3 minutes before the start of increase in the intensities. Moreover, the Doppler shifts started to increase about a minute later than the start of increases in the line widths. 

During the event, the spectral lines of \ion{O}{4} and \ion{S}{4} appeared in the spectra, which were not present in the pre-event conditions. This provided us an opportunity to estimate electron densities during the event \citep{2016A&A...594A..64P}. For this purpose, we used the density sensitive line pair of O~{\sc iv} $\lambda$1399/$\lambda$1401. We obtained average line profile at \textquoteleft A\textquoteright\ between the time
interval of 13:03:41~UT to 13:04:40~UT to improve the signal. Figure~\ref{fig:density}, shows the averaged line profiles (left panel) and theoretical intensity ratio curve as function of electron density obtained from CHIANTI 
database \citep[right panel,][]{1997A&AS..125..149D,2015A&A...582A..56D}. \ion{O}{4}~1399 \& 1401~{\AA} lines are found to be blended with other cool lines in the red wings \citep{2015arXiv150905011Y}. Thus,
to obtain the intensity of the two \ion{O}{4} lines, we fitted the average profiles with two Gaussian functions. Cool lines were fitted with narrow Gaussian function with line widths near to thermal and instrumental broadening. The fitted profiles are also over-plotted in the left panel of Figure~\ref{fig:density}. We obtained the intensity ratio of \ion{O}{4} $\lambda$1399/$\lambda$1401 line pair, which corresponds to an electron number density of 10$^{11.55}$~cm$^{-3}$. 

The two ARTBs and the associated hot loop were also observed with EIS using a raster mode. The EIS raster started at 13:02:20~UT and ended at 13:20:23~UT. Although using EIS raster, one can not study the time evolution of the event, various other useful information such as electron density and flows at coronal temperatures of the event can be obtained. As ARTBs, as well as the hot loop, were changing with the time, each EIS raster steps provide different information of the loop and ARTBs at the different time. 

\begin{figure*}[htbp]
\centering
\includegraphics[width=0.95\textwidth]{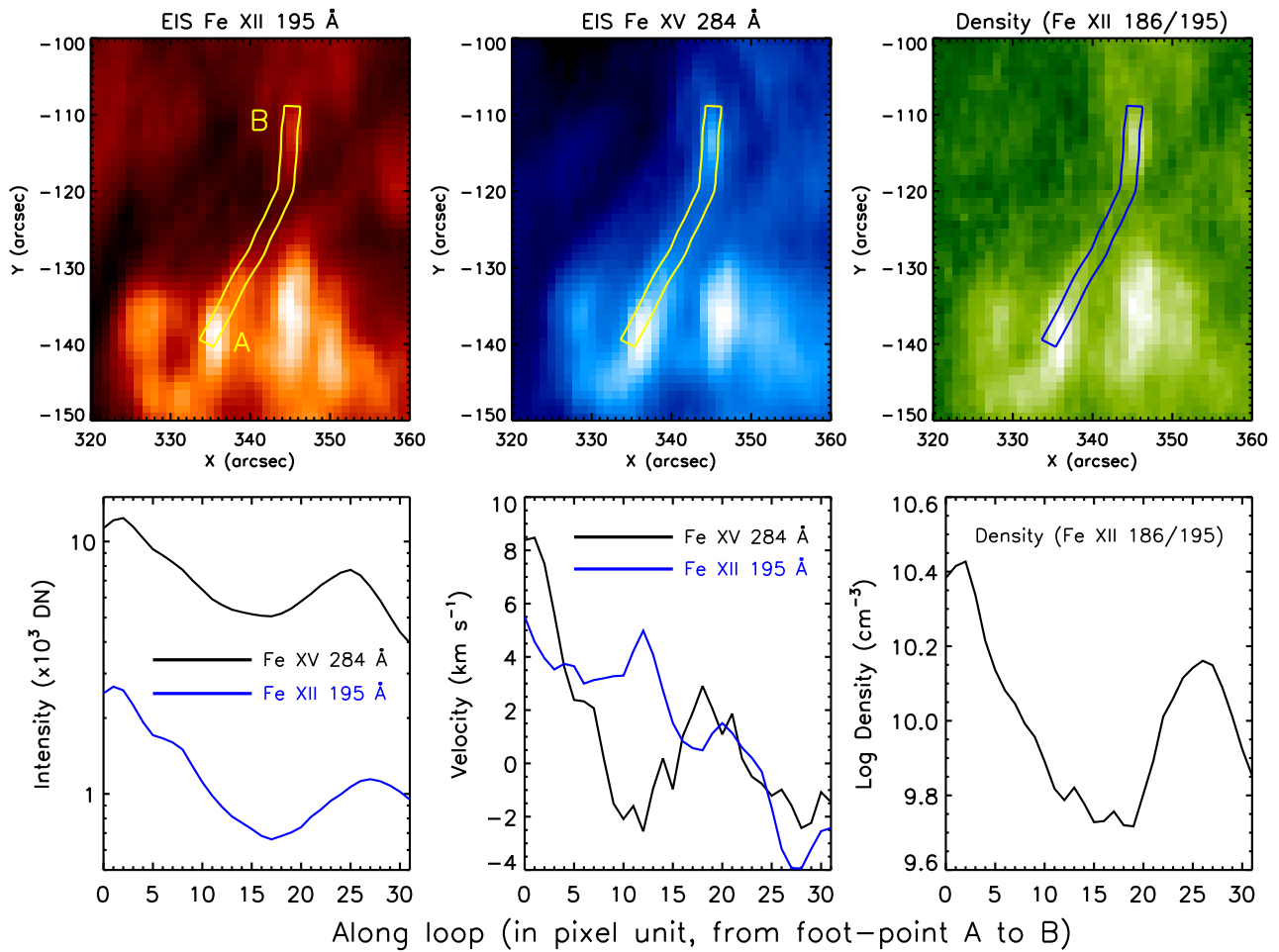}
\caption{EIS intensity maps obtained in \ion{Fe}{12}~195~{\AA} (top left), and \ion{Fe}{15}~284~{\AA} (top middle) and density map (log $n_e$=8.5 to 10.5) obtained using \ion{Fe}{12}~$\lambda 186/\lambda 195$
line pair (top right) as labeled. The bottom row panels show the variation of intensity, velocity, and electron number density along the loop length starting from \textquoteleft A\textquoteright\ to
\textquoteleft B\textquoteright .}
\label{fig:eis}
\end{figure*}

Figure~\ref{fig:eis} show the intensity maps obtained in \ion{Fe}{12} 195~{\AA} (top left) and \ion{Fe}{15} 284~{\AA} (top middle) lines. The top right panel shows the electron density map obtained using the \ion{Fe}{12} $\lambda 186/\lambda 195$ line pair. As raster was taken during the evolution of loop, different phases of loop and foot-points were observed. This makes identification of different parts of the loop difficult with respect to those observed from AIA filters. Although the loop in EIS observation is not distinct, we attempted to trace the loop again from EIS raster with length equal to that of AIA loop and over-plotted in top panels of Figure~\ref{fig:eis}. The traced loop was rastered between 13:09:33 to 13:13:10~UT. We obtained parameters such as intensity, velocity, and electron number density along the EIS loop and plotted them in bottom panels of Figure~\ref{fig:eis}. Intensity and density at both the foot-points are higher than those at the associated loop. Density estimated at \
\textquoteleft A\textquoteright\ and \textquoteleft B\textquoteright\ are 10$^{10.45}$ and 10$^{10.15}$ cm$^{-3}$ respectively whereas that near loop-top is about 10$^{9.75}$ cm$^{-3}$. Doppler velocities were in the range of {--}4 to {+}8 km~s$^{-1}$, which is within the limits of uncertainty (7{--}8~km~s$^{-1}$) as is suggested by \citet{2012ApJ...744...14Y}.

\subsection{Energetics of ARTBs}
\label{sec:energetics}
Using the AIA, IRIS, and EIS observations, we have obtained several parameters that can be used to estimate the order of magnitude of energy released during the ARTBs. 

Similar to \citet{2014Sci...346C.315P}, the thermal and kinetic energy density can be written as 

\begin{equation}
 	E_{th}=3/2 ~n_e~k_B~T
\end{equation}

\begin{equation}
 	E_{kin}=1/2~n_e~\mu~v^2
\end{equation}

where $n_e$ is electron number density, $k_B$ is the Boltzmann constant, $T$ is temperature, $v$ is flow speed, and $\mu$ is mean molecular weight ($\approx 0.6m_p$ for fully ionized plasma). Thus, the total energy released in volume $V$ will be given by  $V (E_{th}+E_{kin}) $.  

To estimate energy released during the ARTBs, we use parameters obtained from cool transition region lines (\ion{O}{4} and \ion{Si}{4}) during the peak phase of the transient as
$n_e$ $\approx$ 10$^{11.55}$ cm$^{-3}$, $T$ $\approx 1.4\times10^5$ K, $v$~$\approx$ 20~km~s$^{-1}$. We consider the volume $V$ of the emission to be $\approx 1500~\times1500~\times~1500$ km$^3$, corresponding to a region of 2\arcsec cubical structure. Using these values, we find that the total energy released is $\approx 3.7 \times 10^{25}$~erg. We also estimate total energy released at hot temperatures to be $\approx 7 \times 10^{26}$~erg by using the parameters obtained from hot AIA channels as $n_e$ $\approx 9\times 10^{10}$ cm$^{-3}$, $T$ $\approx 10$ MK, $v$~$\approx$ 200~km~s$^{-1}$, and $V$~$\approx 1500~\times1500~\times~1500$ km$^3$. We note here that these are just the lower limit of energy released during ARTBs, as was also pointed out by \citet{2014Sci...346C.315P}. 
Estimated energies are equivalent to the energies involved in a typical micro-flare events.

\section{Comparison with Hydrodynamic Simulations} 
\label{sec:model}

\begin{figure*}[htbp] 
\centering
\includegraphics[width=0.95\textwidth]{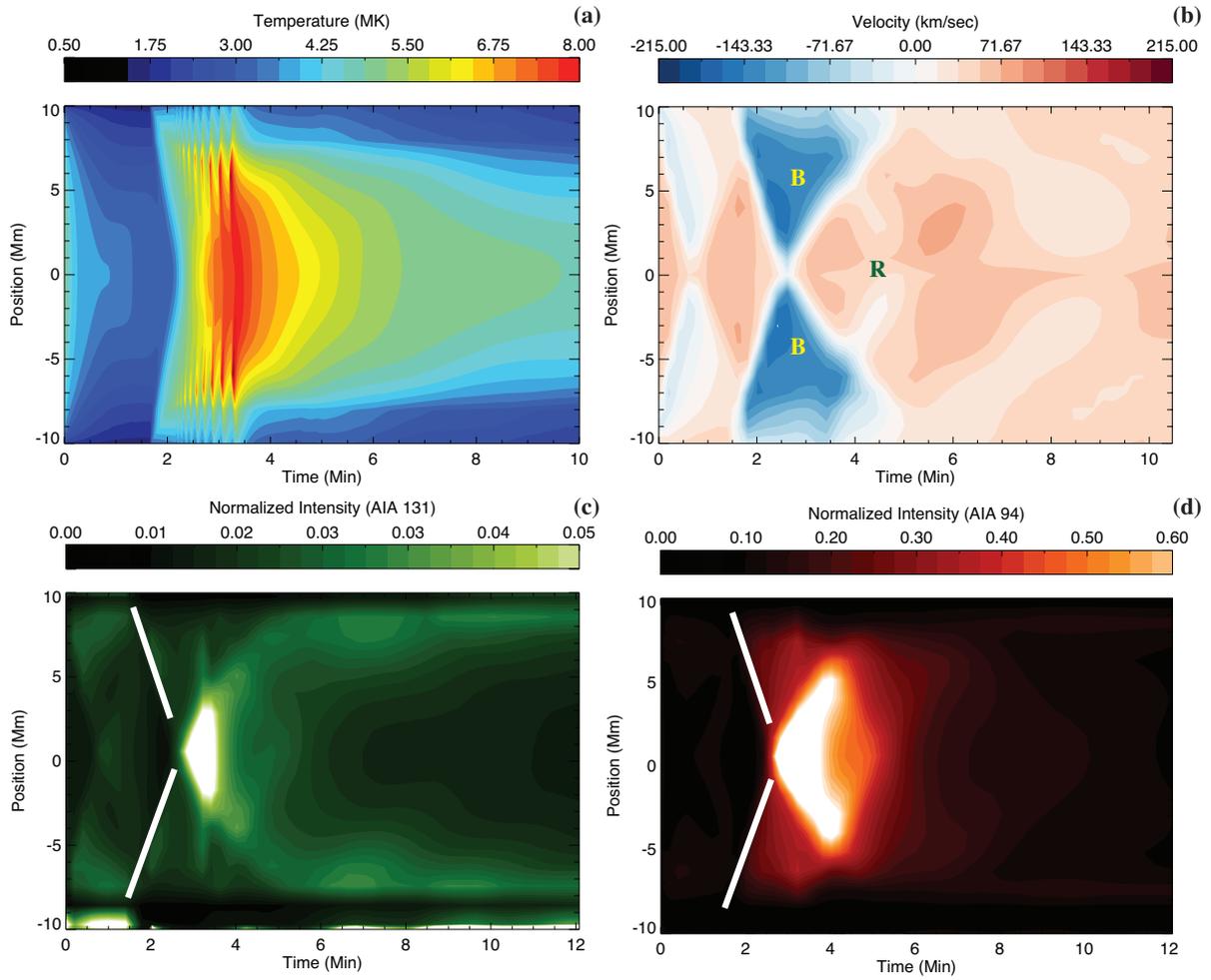}
\caption{The evolution of simulated parameters when heat is deposited at $\pm 9.7$~Mm (a) Temperature evolution along the loop. (b) Velocity evolution along the loop. (c) Synthetic AIA~131~{\AA} emission measure. (d) Synthetic AIA~94~{\AA} emission measure.}
\label{fig:simul_param}
\end{figure*}
\begin{figure*}[htbp] 
\centering
\includegraphics[width=0.95\textwidth]{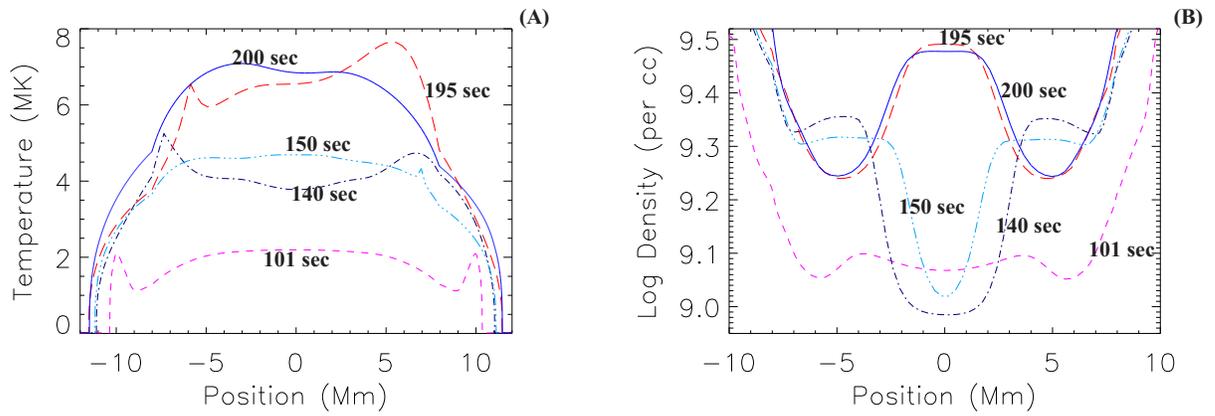}
\caption{Evolution of (A) temperature and (B) density along the simulated loop described in Figures~\ref{fig:simul_param}.} \label{fig:simul_loop}
\end{figure*}

For a better understanding of the observations described in earlier sections, it is instructive to numerically simulate the scenario where a coronal loop may get brightened up following ARTBs.
The observed downflows (shown in Section~\ref{sec:spectroscopy}) in cooler lines (such as  \ion{C}{2} and \ion{Si}{4}) associated with ARTBs suggest a strong chromospheric
condensation due to impulsive heating, similar to those observed in flaring loop \citep[e.g., ][]{2006ApJ...638L.117M,2006SoPh..234...95D,2006SoPh..239..173D} or predicted by \citet{2014Sci...346B.315T}.
In this scenario, heat from the coronal part of the loop gets conducted to the chromosphere, that pushes down the chromospheric plasma further during its initial phase and eventually makes it evaporate into the corona. Based on this assumption, we perform two simulations of a monolithic loop. In one of the simulations, the loop is heated by two identical microflare-like heating events of energy~$10^{27}$~erg, while in the second one 
heat is deposited close to the loop-top via a single microflare-like event. As we demonstrate below, location of heating does not matter much to reproduce qualitative aspects of the observed upflow in the loop.

In order to be consistent with observations, the loop length is taken to be $24$~Mm, with $20$~Mm inside the corona. The loop-top of the simulated loop is situated at the center of the coordinate system, whereas each end of the loop is assumed to be embedded $2$~Mm into the transition region and the chromosphere to serve as a plasma reservoir. The radius of the loop is assumed to be $1$~Mm. Gravity is made to change with height. The gravitational potential has the usual $1/r$ fall off with respect to the center of the Sun. The numerical details of the loop are described in \citet{2009ApJ...699.1480S}.

The initial temperature of the coronal part of the loop is taken to be $4$~MK. Density profile maintains an exponential profile with the least density ($1.46 \times 10^9$~cm$^{-3}$) at the loop-top and the highest density ($2.45 \times 10^{11}$~cm$^{-3}$)  towards the footpoints. We allow the loop to cool down through radiation until $100$ s, when the loop is hit by two similar heating events at the two ends of the footpoints within the
corona ($\pm 9.7$~Mm). These heating events last for $120$~s, span over $200$~km (energy $\sim 10^{27}$~erg), and produce plasma dynamics in the loop. Panels (a) and (b) in Figure~\ref{fig:simul_param} depict temperature and velocity evolutions along the loop. Simulation outputs are folded through AIA response functions of 131~{\AA} and 94~{\AA} filters to produce emission measure outputs. To compare these emission measures with real observations we degrade the spatiotemporal resolution to $1000$~km and $12$~s. The resulting emission measure is depicted in panels (c) and (d) of Figure~\ref{fig:simul_param}. Following the deposition of the heating events, both emission measure plots show a continuous brightening of the coronal loop starting at the footpoints which propagates towards the loop-top. Just as in the real observations (compare Figure~\ref{fig:bright_front}), we also mark these locations with white lines.

Soon after the occurrence of heating events at the both footpoints along coronal loop, local plasma of the loop gets heated up and pushes the transition region downward. This is evident in the temperature plot shown in panel (A) of Figure~\ref{fig:simul_loop}. Heat energy from the event sites quickly flows down to the chromosphere through conduction and leads to the evaporation of chromospheric plasma that fills up the loop as is seen in panel (B) of Figure~\ref{fig:simul_loop}. A strong blueshifted (velocity up to $210$~km s$^{-1}$) region is identified in panel (b) of Figure~\ref{fig:simul_param} (marked as B) for the same reason. Comparing panels (b) with (c) and (d) of Figure~\ref{fig:simul_param}, one can conclude that this evaporated blueshifted plasma dictates the loop brightening from footpoints towards loop-top.  Evaporated plasma from both ends travel towards the loop-top and meet each other, at around $195$~s giving rise to high density at the loop-top and creates loop-top brightening in the simulation, similar to the results of \citet{2016ApJ...823...47S}.
 Following the brightening, overly dense loop-top starts to condensate showing redshifted emission (marked as R in panel (b) of  Figure~\ref{fig:simul_param}). Redshifted plasma (velocity up to $65$~km s$^{-1}$) evacuates the loop and as a consequence the loop starts to fade away. Though we observe such downflows in \ion{Fe}{15} and \ion{Fe}{12} lines, the measured speed (4-8 km s$^{-1}$) is much lower than what is predicted by the simulation. The discrepancy may be due to the projection effect that is not taken into account in the simulation.
 
\begin{figure}[htbp] 
\centering
\includegraphics[width=0.45\textwidth]{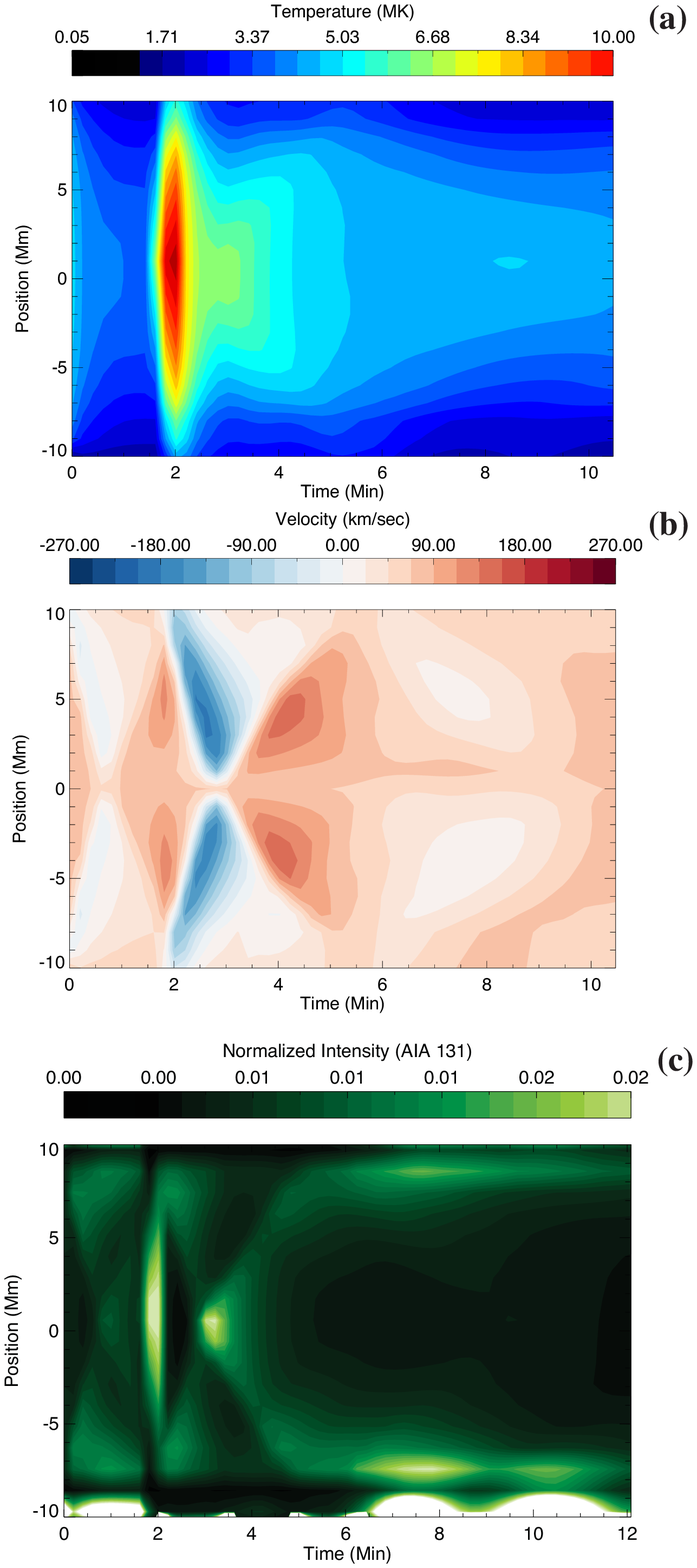}
\caption{The evolution of simulated parameters when heat is deposited close to the loop-top (at $1$~Mm) (a) Temperature evolution along the loop. (b) Velocity evolution along the loop. (c) Synthetic AIA~131~{\AA} emission measure.}
\label{fig:simul_param1}
\end{figure}

Even though we apply heat at the footpoints along coronal loop to reproduce the observed upflow in the loop, the heating location could arguably be anywhere in the coronal part of the loop, as the conduction timescale is much faster than the loop dynamics. Wherever heat is deposited in the corona, heat should get conducted fast to chromosphere to evaporate plasma and should show similar dynamics in the loop. Keeping this in mind we also apply heat close to the loop-top at a single location, $1$~Mm. The amount of heat deposition and other initial conditions remained the same. The result is depicted in panels (a)-(c) of Figure~\ref{fig:simul_param1}. The loop-top heating experiment predicts an initial brightening at the loop-top soon after the deposition of heating event, which is however not in accordance with the present observations. Once heat get conducted to the footpoint a strong upflow (blueshift) is observed, as is evident from panel (b) of Figure~\ref{fig:simul_param1}. These studies therefore 
indicate that the 
location of heating does not matter much to reproduce the observed upflow in the loop. However, evidence of loop-top heating is not clear from the present observations, since there is no evidence of loop-top brightening prior to the dynamics.

\section{Summary and Discussion} 
\label{sec:summary}
In this paper, we have presented detailed study of two simultaneously occurring ARTBs with AIA-HMI/SDO, EIS-XRT/Hinode, and IRIS. Both the ARTBs were observed in all the passbands of AIA, IRIS and XRT, and attained their peak intensity within 36~s and 45~s in AIA and IRIS passbands, respectively. In the hot channels of AIA such as 131~{\AA}, corrected 94~{\AA}, and 335~{\AA} an associated loop was seen that was rooted in the two ARTBs. The full loop became visible following the ARTBs and appeared after $\approx 25$~s in 131~{\AA}, $\approx 40$~s in 94~{\AA}, and $\approx 6.5$~min in 335~{\AA}. Using time-distance plot, we detected the flow of hot plasma from the footpoints towards the loop-top. The estimated flow speeds were found to be larger for high-temperature plasma and smaller for relatively cooler plasma.

To understand the plasma temperature of the hot loop as well as of the two ARTBs, we performed DEM analysis using the observations taken in all the six AIA coronal channels. The DEM analysis suggests that the loop, as well as footpoints, get heated up to 10~MK. Signatures of such a hot plasma are also clearly visible in the XRT images. Electron number density at ARTB (\textquoteleft A\textquoteright) was found to be about $\approx 9\times 10^{10}$ cm$^{-3}$ from DEM analysis.

The spectroscopic study using the IRIS instrument that only passed through one of the ARTBs (\textquoteleft A\textquoteright), revealed strong redshifts in both the lines of \ion{C}{2} and \ion{Si}{4}. It further revealed that the Doppler shifts at the locations of ARTB started to increase a couple of minutes earlier than the intensity. The line width started to increase even earlier. To the best of our knowledge, such observation has not been reported earlier. Such observations may hold the key for the initiation of such events. Using the density sensitive \ion{O}{4} $\lambda$1399/$\lambda$1401 line pair observed with IRIS, we estimated average electron number density during the ARTB to be $\approx 10^{11.55}$ cm$^{-3}$.

EIS raster which scanned the region of interest during 13:09:33 to 13:13:10~UT, showed density at both the ARTBs \textquoteleft A\textquoteright\  and \textquoteleft B\textquoteright\  to be 10$^{10.45}$ and 10$^{10.15}$ cm$^{-3}$ respectively, whereas that near the loop-top is about 10$^{9.75}$ cm$^{-3}$. The obtained density and temperatures from the observations suggest that the total energy (kinetic and thermal) released during the transient is of the order of $\approx 10^{26}$ erg. This is similar to the energies involved in micro-flaring events. 

The observations of downflows in cooler lines like \ion{C}{2} and \ion{Si}{4} and upflows in hot AIA channels during the ARTBs can be understood by invoking the concept of standard flare model. The upflows in the hot lines can be attributed to chromosphere evaporation whereas the downflows in cool lines suggest chromospheric condensation  \citep[e.g.,][and references therein]{2014ApJ...795...10L,2017ApJ...841L...9L}. 
Figure~\ref{fig:bright_front} shows that the hot loop was visible for a time interval of 2{--}4~min,  2{--}9~min, and 2{--}24~min (not shown for full range) in AIA 131, corrected 94, and 335~{\AA} passbands, respectively. This finding suggests that the lifetime of observed hot loop is temperature dependent and that the high temperature upflows are short lived events.

To compare the lifetime  of observed loop at different temperature with expected cooling time due to various loss rates, we estimate combined cooling time due to conduction and radiation as well as chromospheric evaporation given by \citet{1995ApJ...439.1034C}, 

\begin{equation}
 \tau_{cool} \approx  2.35 \times 10^{-2} \frac{L_0^{5/6}}{T_e^{1/6}n_e^{1/6}}
 \label{eq:tcool}
\end{equation}

Using half loop length $L_0 \approx 11\times 10^8$ cm,  temperature $T_e\approx 1\times 10^7$ K, and density as estimated from EIS $n_e \approx 10^{10}$ cm$^{-3}$, we get $\tau_{cool} \approx 20$ min.  This cooling time matches well with the life time of loop observed from AIA 335~{\AA} passband. 

Following \citet{1999ApJ...515..842A}, we further obtain conductive and radiative cooling time as

\begin{equation}
 \tau_{cond} \approx 1.1\times 10^{-9} n_e T_e^{-5/2} L_0^2 [s]
\end{equation}

\begin{equation}
\tau_{rad} \approx \frac{3 n_e k_b T_e}{n_e^2 \varLambda(T_e) }
\end{equation}

where $\varLambda(T_e) \approx 10^{-21.94}$ erg cm$^{-3}$ s$^{-1}$ for EUV loops. Using above mentioned observed parameters, we found $\tau_{cond} \approx 42$~s whereas $ \tau_{rad} \approx  3607$~s. For a temperature around 7~MK, $\tau_{cond}$ takes the values of about 100~s whereas  $ \tau_{rad} $ takes the value of about 2554~s. A comparison of the lifetime of the loop with radiative and conductive cooling time suggests that the conduction is the dominant cooling mechanism for the loop in AIA 131~{\AA} passband. The finding that the conductive losses are much more efficient suggest that there is more energy involved in the events than that being seen as was also suggested by \citet{2014Sci...346C.315P}. Similar results were obtained for Hi-C EUV bright dots studied by \cite{Sub_KT}. 

To better understand the physical mechanism responsible for the observed plasma dynamics, we performed 1-D hydrodynamic simulation of a monolithic loop. In this simulation, energy was deposited at the two footpoints mimicking the two observed ARTBs. The obtained density and temperature from simulations were convolved with the temperature response function of AIA 131~{\AA} and 94~{\AA} channels to produce intensities that would be seen in respective channels. The simulated velocity (panel (b) in Figure~\ref{fig:simul_param}) and forward modeled intensities in 131 and 94~{\AA} channels (panels (c) and (d) in Figure~\ref{fig:simul_param}) closely represent the observations qualitatively. However, we see presence of downflows in the simulated velocities after the collision of upflows, although some downflows are present in our observations but those are inconclusive due to limitations on the velocity measurements with EIS. Very similar dynamics is also observed when the loop is heated close to its apex, except a 
loop-top brightening at the time of the heating event (Figure~\ref{fig:simul_param1}).  

The enhanced emission measure that is observed at the loop-top in forward modeled intensities, which can be 
explained by the collision of the chromospheric evaporation flows \citep[see e.g.,][]{2016ApJ...823...47S}, is also not present  in the observation. Clearly, further observations and modelings are required to fully comprehend the dynamics of plasma in such loop structures.

Understanding of such ARTBs and their effects on the mass and energy transfer in coronal loops is important from the point of view that it may hold the key to coronal heating. The current observation shows that the conduction is the dominant mechanism for cooling. This is suggestive of the fact that there is a large amount of energy involved in such events, which is conducted away, over and above the measured radiative energy 10$^{26}$ ergs. Such energy may even be non-thermal in nature and could be probed with high resolution hard X-ray instruments.

\acknowledgments
GRG is supported through the INSPIRE Faculty Award of the Department of Science and Technology (DST), India. DT acknowledges the Max-Planck India Partner Group on Coupling and Dynamics of the Solar Atmosphere at IUCAA.  AIA and HMI data are courtesy of SDO (NASA). IRIS is a NASA small explorer mission developed and operated by LMSAL with mission operations executed at NASA Ames Research center and major contributions to downlink communications funded by the Norwegian Space Center (NSC, Norway) through an ESA PRODEX contract. Hinode is a Japanese mission developed and launched by ISAS/JAXA, collaborating with NAOJ as a domestic partner, NASA and STFC (UK) as international partners. Scientific operation of the Hinode mission is conducted by the Hinode science team organized at ISAS/JAXA. This team mainly consists of scientists from institutes in the partner countries. Support for the post-launch operation is provided by JAXA and NAOJ (Japan), STFC (U.K.), NASA (U.S.A.), ESA, and NSC (Norway). Facilities: SDO (
AIA, HMI). CHIANTI is a collaborative project involving George Mason University, the University of Michigan (USA) and the University of Cambridge (UK).


\end{document}